\documentclass[12pt]{article}
\usepackage{graphicx,amsmath,amssymb,amsfonts,color}
\usepackage{citesort}

\input{text/StrAsym.pre}
\input{text/StrAsym.fig}

\begin{document}


\begin{titlepage}

\begin{tabular}{l}
\noindent\DATE
\end{tabular}
\hfill
\begin{tabular}{l}
\PPrtNo
\end{tabular}

\vspace{1cm}

\begin{center}
\renewcommand{\thefootnote}{\fnsymbol{footnote}}
{ \LARGE \TITLE }                           

\vspace{1.25cm}
{\large  \AUTHORS}          
\vspace{1.25cm}

\INST                          
\end{center}

\vfill
\ABSTRACT                 
\vfill

\end{titlepage}

\renewcommand{\thefootnote}{\alph{footnote}}   
\setcounter{footnote}{0}



\tableofcontents  \newpage

\section{Introduction}

The recent measurements of both neutrino and antineutrino production of dimuon
final states (charm signal) by the CCFR and NuTeV collaborations \cite{CN}
provide the first promising direct experimental constraints on the strange and
anti-strange quark distributions of the nucleon, $s(x)$ and $\bar{s}(x)$. In
addition to the intrinsic interest in nucleon structure \cite{models},
the strange asymmetry
($s-\bar{s}$) has important implications on the precision measurement of the
Weinberg angle in deep inelastic scattering
of neutrinos.\cite{NuTeV,BPZ,forte,moch,gambino,paper2}
We report here the first global QCD analysis that
includes the new dimuon data, using the methods developed by the
CTEQ collaboration, specifically to explore the strange and anti-strange parton
parameter space.%
\footnote{A preliminary version of this study was reported at the Lepton Photon
2003 International Symposium (LP2003), Fermilab, August 2003.  Cf.~P.~Gambino
\cite{Gambino:2003xc} and R.~Thorne \cite{Thorne:2003gu}, published in
the Proceedings of LP2003.}%

In previous global analyses, information on $s$ and $\bar{s}$ has resided only
in inclusive cross sections for neutral and charged current DIS. The
reliability of the extraction of the quite small $s$ and $\bar{s}$
components
\footnote{%
The strangeness content of the nucleon, as measured by the
momentum fraction carried by $s$ or $\overline{s}$,
is of order 3\% at $Q=1.5$ GeV.} %
(from differences of large cross sections measured in different experiments)
was always in considerable doubt.
For this reason, most global fits adopted the assumption
$s(x)=\bar{s}(x) = \kappa(\bar{u}+\bar{d})/2$;
this approximation was inferred from the earlier combined neutrino
and antineutrino dimuon experiments \cite{dmdata}
which extract a consistent value of
\begin{equation}\label{eq:kappa}
\kappa \equiv \frac{\int dx\ x \left[s(x,Q^2) + {\bar s}(x,Q^2)\right]}
{\int dx\ x \left[{\bar u}(x,Q^2) + {\bar d}(x,Q^2)\right]}
\sim 0.4
\end{equation}
at some low value of $Q$.
The recent high-statistics dimuon
measurements of \cite{CN} provide greater accuracy,
and the high purity of separate neutrino and anti-neutrino events
offers the first
opportunity to study the difference $s(x)-\bar{s}(x)$.
Neutrino induced dimuon
production, $(\nu/\bar{\nu}) N \rightarrow \mu^+ \mu^- X$, proceeds primarily
through the subprocesses
$W^+s \to c$ and $W^-\bar{s} \to \bar{c}$ respectively,
and hence provides independent information on $s$ and $\bar{s}$.

We present the first global QCD analysis that includes this new dimuon data.
The new results demonstrate, first of all, that the strangeness
asymmetry, as measured by the momentum integral
\begin{equation}
[S^{-}]\equiv \int_0^1 x [s(x)-\bar{s}(x)] dx\ \ \ ,
\label{eq:Sm}
\end{equation}
is indeed more sensitive to the dimuon data than to the other
DIS data.
We then use the recently developed Lagrange multiplier
method of global analysis to explore the range of uncertainty of
$[S^-]$.
In this first report, we concentrate
on $[S^-]$, the integrated strangeness asymmetry, which represents a new
parton degree of freedom in the nucleon heretofore largely unexplored, and
which has immediate impact on precision electroweak physics because of the
NuTeV anomaly.\cite{NuTeV} A full exploration of the density functions
$s(x)$ and $\bar{s}(x)$ will be presented later.

We begin by describing the general features of the strangeness sector of the
nucleon structure in the QCD framework, and our general parametrization of
that sector. A brief review and discussion of the pQCD calculations
that are relevant to the interpretation of the dimuon data
is then followed by
the main results of the global analysis, with
emphasis on concrete representative global fits relevant for probing the
strangeness asymmetry, and on Lagrange multiplier results for the
integrated momentum asymmetry (\ref{eq:Sm}).
The paper concludes with a summary of the extensive
studies performed beyond the examples given, comparisons to previous work on
strangeness asymmetry, and conclusions.

\section{General properties of $s(x)-\bar{s}(x)$ and its first two moments}
\label{sec:general}

Before we discuss the concrete fitting procedure and the results
in the following Sections,
we feel it is instructive to formulate the qualitative expectations
that are based on general QCD requirements (before any data).

\subsection{Requirements}

For each $Q,$ let us define the strangeness \emph{number densities}
$s^{\pm }(x)$ and their integrals $[s^{^{\pm }}]$ by%
\begin{equation}
[s^\pm]\equiv \int_{0}^{1}\,s^\pm(x)\,dx\equiv  
\int_{0}^{1}\,[s(x)\pm\bar{s}(x)]\,dx  \; ,     
\label{eq:StrNo}
\end{equation}%
and the \emph{momentum densities} $S^{\pm }(x)$ and integrals $[S^{\pm }]$ by%
\begin{equation}
[S^{\pm }]\equiv \int_{0}^{1}S^{\pm }(x)\,dx \equiv 
\int_{0}^{1}x[s(x)\pm \bar{s}(x)]\,dx \; . 
\label{eq:StrMom}
\end{equation}
In the QCD parton model, certain features of these quantities are necessary:
\begin{Simlis}{1em}
\item The parton distributions $s(x,Q^2)$ and $\bar{s}(x,Q^2)$\footnote{As we
had done already in Eqs.~(\ref{eq:StrNo}), (\ref{eq:StrMom}) above, we will
suppress the obvious $Q^2$ dependence from now on.} [or equivalently
$s^{\pm}(x)$], are parametrized at some low (but still perturbative) scale
$Q_{0}$; the full $Q$-dependence is then determined by DGLAP
evolution.\footnote{For definiteness, the CTEQ PDFs evolve from $Q_0=m_c=1.3\
{\rm GeV}$.}

\item The strangeness number sum rule for the nucleon requires
\begin{equation}\label{eq:NumSR}
[s^-]=0 \qquad \mbox{(for all $Q$)}.
\end{equation}
A necessary corollary is that the density function $s^{-}(x)$ must be less
singular than $1/x$ as $x\rightarrow 0$ for all $Q$.
As DGLAP evolution
preserves Eq.~(\ref{eq:NumSR}) (which is a consequence of the
conservation of the strange vector current
$J^{\mu}_s = {\bar \psi}_s\gamma^{\mu} \psi_s$),
it suffices to impose it at $Q_0$.

\item The momentum sum rule requires
\begin{equation}\label{eq:MomSR}
[S^+] = 1 - \Sigma_0 \qquad \mbox{(for all $Q$)},
\end{equation}
where $\Sigma_0$ represents the momentum fraction of all non-strange partons.
Through this condition the global inclusive DIS and other
data---which directly affect $\Sigma_0$---indirectly also constrain $[S^+]$.

\item In the limit $x\rightarrow 0$ (high energy and fixed $Q$), Regge
considerations and the Pomeranchuk theorem predict
$s^{-}(x)/s^{+}(x)\rightarrow 0$.

\end{Simlis}

\subsection{Expectations}

From the above general constraints, we draw the following conclusions:

\begin{Simlis}{1em}
\item[(i)] The number sum rule, Eq.~(\ref{eq:NumSR}), implies that a graph
of $s^{-}(x)$ must cross the $x$-axis {\em at least once} in the interval
$0 < x < 1$;
and the areas bounded by the curve above and below the $x$-axis must
be equal.

\item[(ii)]
Assuming a simple scenario in which there
is---as supported by theoretical models \cite{models}
based on Lambda-Kaon fluctuations---only
one zero crossing, either
$s^{-}(x)<0$ in the low $x$ region and $s^{-}(x)>0$ in the high $x$
region, or vice versa.  The two possibilities imply $[S^{-}]>0$ or $[S^{-}]<0$,
respectively, because the momentum integral suppresses the small $x$ region and
enhances the large $x$ region.

\item[(iii)] The low-$x$ behaviour of
$\left| s^{-}(x) \right| \sim x^{\beta_{-}}$
will then be correlated with the size of $\left| [S^-]\right|$: The steeper
the function $s^{-}(x)$, the larger $\left| S^{-}(x) \right|$ has to be at
large $x$.

\end{Simlis}
\figA
To illustrate point (ii), we preview
a representative CTEQ fit
(to be discussed in full detail in Sec.~\ref{sec:global}) in
Fig.~\ref{fig:A} and juxtapose it with results from previous literature.
According to the parametrizations of $s(x)$ and $\bar{s}(x)$ used by the
CCFR-NuTeV dimuon study \cite{CN,NuTeV2}, $s^{-}(x)$ is negative in the $x$
range covered by the experiment ($0.01 < x < 0.3$).
\footnote{Since the preliminary version of this work was reported, the
CCFR-NuTeV collaboration has emphasized that their most recent analysis 
favors an integrated strangeness asymmetry
that is consistent with zero
(K.~MacFarland and P.~Spentzouris, communications at
the LP03 Symposium and WIN03 Workshop). More definitive studies are needed to
clarify the situation. See further discussions in Sec.~\ref{sec:compare}.} 
A previous detailed global analysis of inclusive data by Barone et
al.~\cite{BPZ} (BPZ), finds that $s^{-}(x)$ is
positive in the large $x$ region. These previous results are shown in
Fig.~\ref{fig:A} as the dot-dashed and dashed curves, respectively.
In light of the theoretical constraints discussed above, both these results
would hint at
the first possibility mentioned above, i.e.~$[S^{-}]>0$.  This is not
the case, however, for the CCFR-NuTeV curve, as it does not cross the
$x$-axis; it violates the strangeness number sum rule and cannot be
taken to represent the physics outside a limited window in $x$.  The
BPZ curve does satisfy the sum rule; we note (Fig.~\ref{fig:A}) it has
two zero-crossings.
The solid curves (Class B) in the
two plots of Fig.~\ref{fig:A} provide a concrete example of
$s^-(x)$ and $S^-(x)$ following the above requirements and expectations.

To illustrate point (iii), we preview
various classes of solutions (again, to be discussed in detail
below) in the upper plot of Fig.~\ref{fig:B}.
Because the experimental constraints are weak or non-existent in the
very small $x$ region, say $x<0.01$,
the detailed behavior of $s^{-}(x)$ is unconstrained in
this region. However, this uncertainty at small $x$ is
considerably reduced in $S^-(x)$, as demonstrated by the curves
of the lower plot.
Thus the above observations concerning $[S^{-}]$
are affected only mildly by the uncertainty of the very
small $x$ behavior, unless that behavior is so extreme
that the small $x$ region provides a significant contribution
to the number sum rule. We will refrain from
exploiting such a mathematical possibility as long as it does not seem to be
motivated by any physics.\footnote{A residual bias of the results
on the assumed functional forms, or rather on the rejection of some forms
as merely mathematical and not physical, is unavoidable in parton
analysis. 
One can imagine the extreme (unphysical)  scenario of spike-like structures
escaping ``detection'' in the $x\rightarrow 0$ region but, nevertheless,
affecting sum rules.
}


\figB


\section{General parametrization of the strangeness distributions}
\label{sec:param}

To explore the strangeness sector of the parton structure of the nucleon,
we need a suitable parametrization of $s(x)$ and $\bar{s}(x)$
(or equivalently $s^{\pm}(x)$) at a fixed scale $Q_{0}$.
This parametrization must satisfy the
theoretical requirements specified above, and it should be as general as
possible so that the allowed functional space can be fully explored.  A general
form is essential, so that our conclusions are not artifacts of the
parametrization, but truly reflect the experimental and theoretical constraints.
In the following we explain our choice of such a parametrization. Full
details including explicit parameters are given in an Appendix.

It is more natural to parametrize the $s^{\pm}(x,Q_0)$ functions
independently (rather than $s$ and $\bar{s}$) since they satisfy different QCD
evolution equations: pure non-singlet for $s^-$ and mixed
singlet/non-singlet for $s^+$. We use the following parametrizations,
\begin{eqnarray}
s^{+}(x,Q_{0}) &=&A_{0}\,x^{A_{1}}(1-x)^{A_{2}}P_{+}(x;A_{3},A_{4},...) %
\label{eq:StrParam1} \\
s^{-}(x,Q_{0}) &=&s^{+}(x,Q_{0}) \tanh [a\,x^{b}(1-x)^{c}P_{-}(x;x_0,d,e,...)] %
\label{eq:StrParam2}
\end{eqnarray}%
where $P_{+}(x;A_{3},\dots)$ is a positive definite, smooth function in the
interval $(0,1)$, depending on additional parameters {$A_{3},\dots$} such
as are used for $u,d,g,\dots$ in most CTEQ \cite{cteq6m} and other global
analyses \cite{mrst,grv98}; and
\begin{equation}\label{eq:StrParam3}
P_{-}(x)=\left(1-\frac{x}{x_{0}}\right)
\left(1 + dx+ex^{2}+\cdots \right)
\end{equation}%
where the crossing point $x_{0}$ is determined by the strangeness number sum
rule $[s^{-}]=0$, and the parameters {$d,e,\dots$} are optional, depending on
how much detail is accessible with the existing constraints. Important
features of this parametrization are the following:
\begin{Simlis}{1em}
\item The strangeness quantum number sum rule, $[s^{-}]=0$, is satisfied
by the choice of $x_0$. The parameter $x_{0}$ has a physical interpretation:
it is the ``crossing point'' where $s^-(x)=0$.  (If $d$ and/or $e$ are not
zero, there can be additional zeros of $s^-(x)$; in practice---as explained
in the previous section and below in Sec.~\ref{sec:procedure}---we restrict
our analysis to solutions with a single crossing only.)

\item The fact that the $\tanh $ function has absolute value less than 1
ensures positivity of $s(x)$ and $\bar{s}(x)$. The fact that $\tanh $ is a
monotonic function guarantees that the function $s^{-}(x)$ can be made as
general as necessary by the choice of $P_{-}(x)$.

\item The small-$x$ behavior of $s^{-}(x)$ must be such that the integral
$[s^{-}]$ converges (before the root $x_{0}$ is determined). Let
$\beta_-\equiv A_{1}+b$; then Eq.\,(\ref{eq:StrParam2}) implies
\begin{equation}
s^{-}(x)\sim x^{\beta_-} \  \mathrm{as} \ \ x\rightarrow 0.
\label{eq:betam}
\end{equation}
The convergence of $[s^{-}]$ is guaranteed if $\beta_- > -1$, i.e.~the
parameter $b$ is chosen in the range $b>-1-A_{1}$.
\end{Simlis}
Because $P_{\pm }(x)$ can be made as general as necessary, the
choice in Eqs.\,(\ref{eq:StrParam1})--(\ref{eq:StrParam3}) is capable of
exploring the full strangeness parameter space allowed by data in the PQCD
framework.

Detailed formulas for all flavors used in the actual analysis are presented
in the Appendix.



\section{Global Analysis}

\label{sec:global}

We now describe the global QCD analysis, which includes all relevant
experimental data and implements the theoretical ideas outlined above. This
may be considered an extension of the on-going CTEQ program of global
analysis. Several new elements (compared to the latest CTEQ6M \cite{cteq6m}
analysis) are present. On the experimental side, we have added the CDHSW
inclusive $F_{2}$ and $F_{3}$ data sets \cite{cdhsw}, and the CCFR-NuTeV
dimuon data sets \cite{CN}. On the theoretical side, we have expanded the
parameter space to include the strangeness sector as discussed in Sec.~\ref%
{sec:param}.

Compared to the global analyses of BPZ \cite{BPZ}, which also allow $s\neq
\bar{s}$, the major difference experimentally is our inclusion of the dimuon
data, which provide a direct handle on $s$ and $\bar{s}$; and,
theoretically, the generality and naturalness of our parametrization of the
strange distributions.
\footnote{%
Ref.\,\cite{BPZ} parametrizes $s(x)$ and $\bar{s}(x)$
rather than $s^\pm(x)$.} Since the results of
\cite{BPZ} on strangeness asymmetry rely on small differences of
inclusive DIS charged-current and neutral-current measurements, BPZ
performed the analysis at the cross section level, applying uniform
procedures to treat data from different experiments in the comparison
to theory.  Considering small differences between inclusive cross
sections, the possible strange asymmetry is but one of many sources
that could lead to such differences.

As the dimuon data more directly constrain the strange PDFs, this is
an important new element to our fit.  For all inclusive DIS processes
we use the standard procedure of comparing theory with the published
$F_2$ and $F_3$ structure function data. In our analysis, the fit to
charged-current (neutrino) inclusive structure functions is dominated
by the high statistics CCFR data. Although we have included the
earlier inclusive CDHSW data (which play a prominent role in the
analysis of \cite{BPZ}), they have no discernible influence on the
results presented below.

To include the CCFR-NuTeV neutrino and antineutrino dimuon production data
in a global QCD analysis is not a straightforward task. The experimental
measurement is presented as a series of \textquotedblleft forward
differential cross sections\textquotedblright\ with kinematic cuts, whereas
the theoretical quantities that are most directly related to the parton
distribution analysis are the underlying (semi-) inclusive \textquotedblleft
charm quark production cross sections\textquotedblright .\ The gap between
the two is bridged using a Monte Carlo program that incorporates kinematic
cuts as well as fragmentation and decay models. In our analysis, we use a
Pythia program provided by the CCFR-NuTeV collaboration to do this
efficiency-correction.\footnote{%
We thank Tim Bolton and Max Goncharov, in particular, for providing this
program, as well as assistance in its use. Their help was vital for carrying
out this project.\label{fn:Bolton}} 
This Monte Carlo calculation is done in the spirit and the framework of
leading-order (LO) QCD. CTEQ5L parton distributions and Peterson
fragmentation functions were used. The parameters of the model were tuned to
reproduce, as closely as possible, the detailed differential dimuon cross
sections published in \cite{CN}.

\subsection{$d\protect\sigma ^{\protect\nu N\rightarrow cX}$ in QCD}

At LO in PQCD, the cross section formula for $\nu N\rightarrow cX$ is \cite%
{gkr1}
\begin{eqnarray}
\xi s^{\prime }(\xi ,Q^{2})_{\mathrm{eff}}\  &\equiv &\frac{1}{2}\ \frac{\pi
(1+Q^{2}/M_{W}^{2})^{2}}{G_{F}^{2}M_{N}E_{\nu }}\ \left\vert
V_{cs}\right\vert ^{-2}\ \frac{d^{2}\sigma ^{\nu N\rightarrow cX}}{dx\ dy}
\notag \\
&=&\ (1-\frac{m_{c}^{2}}{2M_{N}E_{\nu }\xi })\ \xi s^{\prime }(\xi ,Q^2)
+{\mathcal{O}}(\alpha _{s})\ \ \ ,  \label{eq:seff}
\end{eqnarray}%
with the CKM matrix element $\left\vert V_{cs}\right\vert $ and where the
Barnett-Gottschalk parameter \cite{barnett,gotts}
\begin{equation}
\xi \equiv x\ \left( 1+\frac{m_{c}^{2}}{Q^{2}}\right)
\end{equation}%
approaches Bjorken-$x$ as $Q\rightarrow \infty $ (relative to
$m_{c}=1.3\ \mathrm{GeV}$). The quantity $s^{\prime }(\xi ,Q^{2})_{%
\mathrm{eff}}$ in Eq.~(\ref{eq:seff}) includes Cabibbo suppressed
contributions in neutrino scattering via
\begin{equation}
s^{\prime }\equiv s+\frac{\left\vert V_{cs}\right\vert ^{2}}{\left\vert
V_{cd}\right\vert ^{2}}\ d
\end{equation}%
with obvious adjustments for the anti-neutrino case\footnote{$q\rightarrow {%
\bar{q}}$ for $q=s,d$.}.

The NLO corrections to $\xi s^{\prime }(\xi ,Q^{2})_{\mathrm{eff}}$, defined
via the perturbative series
\begin{equation}
\xi s^{\prime }(\xi ,Q^{2})_{\mathrm{eff}}=\sum_{i}\alpha _{s}^{i}\ \xi
s^{\prime }(\xi ,Q^{2})_{\mathrm{eff}}^{(i)}\ \ \ ,
\end{equation}%
are generically of the form
\begin{equation}
\xi s^{\prime }(\xi ,Q^{2})_{\mathrm{eff}}^{(1)}\propto \sum_{f=g,s^{\prime
}}\ f\otimes H_{f}  \label{eq:nlo}
\end{equation}%
with $\otimes $ denoting a convolution integral over parton momentum. These
were first calculated more than 20 years ago \cite{gotts}. Later
calculations \cite{gkr1,ks} corrected minor typos
and employed the modern ${%
\overline{\mathrm{MS}}}$ renormalization scheme and the ACOT treatment \cite%
{acotcc} of amplitudes with massive quarks ($m_{s,c}\neq 0$). Very recently,
the NLO charm production contributions to the full set of electroweak
structure functions were calculated \cite{kr}, including terms that are
suppressed by $m_{\mu }^{2}/ME_{\nu }$. In order to apply detector
acceptance corrections to the data \cite{CN}, differential NLO distributions
were calculated in \cite{gkr2} and \cite{disco} that provide the charm
hadron (D meson) kinematics in terms of the fragmentation variable $z$ and
rapidity $\eta $. The $d\sigma /dxdydzd\eta $ code DISCO \cite{disco},
written by two of the authors of the present article in collaboration with
D.~Mason of NuTeV, exists as an interface to the NuTeV MC event generator.
Detailed results can be found in the articles listed above. It suffices to
say: (i) the NLO calculations all agree; and (ii) for the fixed target
kinematics under investigation, the NLO corrections to the LO results are
modest---no bigger than $\lesssim 20\%$ (see Fig.~1 in \cite{gkr1}).

As mentioned before, the global fits performed in our study are extensions
of  the full NLO CTEQ6 analysis with the addition of constraints due to
neutrino dimuon production. For the latter process, we have done extensive
studies using either the LO formula, Eq.~(\ref{eq:seff}), or the NLO
treatment of \cite{gkr1}, Eq.~(\ref{eq:nlo}). The results obtained in the
two cases are quite similar. For definiteness, the main results presented in
Sec.~\ref{sec:results} are those obtained by using Eq.~(\ref{eq:seff}),
since the acceptance corrections made to the data set are currently based on
a LO model, as mentioned earlier. \ Since we have determined that the NLO
corrections to the hard cross section are small (compared to experimental
errors, for instance), and since we found the uncertainty range of the main
result (on $[S^{-}]$ ) is much broader than the difference between the
central values obtained by using Eq.~(\ref{eq:seff}) with or without
the corrections in Eq.~(\ref{eq:nlo})
(cf.~Sec.~\ref{sec:compare}), this approximation does not affect the outcome
of our analysis.\footnote{%
It is certainly desirable to have the inclusive cross sections corrected for
acceptance based on NLO models (such as \cite{disco}), that can be compared
to (\ref{eq:nlo}) in a full NLO global analysis.\ This is under active
development by a theory (CTEQ)-experiment (CCFR-NuTeV) collaboration.}

\subsection{Procedure}

\label{sec:procedure}

Our analysis is carried out in several stages. First we must find
appropriate starting values for the fitting parameters.
For this purpose, we implement the following steps.

\begin{Simlis}{1em}
\item We rerun the CTEQ6M global fit with the added CDHSW inclusive neutrino
scattering data, keeping all other conditions the same. \ This intermediate
fit is extremely close to the CTEQ6M one, since the fit to inclusive DIS
data is totally dominated by the high statistics neutral current experiments
on the one hand, and the CCFR charged current experiment on the other.

\item We then fix all of the \textquotedblleft
conventional\textquotedblright\ parton parameters to their values in this
intermediate fit, and fit the complete set of data, including the new dimuon
data, by varying only the parameters associated with the new degrees of
freedom in $s^{-}$. We obtain results consistent with expectations:

\begin{Simlis}{1em}
\item[(i)] Most of the data sets used in the previous analysis are not
affected at all by the variation in $s^{-}$.

\item[(ii)] A few fully inclusive cross sections are slightly affected by
the variation of $s^{-}$, mainly:

\begin{Simlis}[]{1em}
\item[a.] $F_{3}$ which depends on $u-\bar{u}+d-\bar{d}+s-\bar{s}\,\dots $.

\item[b.] The $W^\pm$ charge asymmetry which receives a contributions from
$gs \rightarrow W^- c$.
\end{Simlis}

These sensitivities to $s^{-}$ are weak.

\item[(iii)] The CCFR-NuTeV dimuon data sets are the most constraining ones
for fitting $s^{-}(x)$.
\end{Simlis}

We obtain good fits using either the 3-parameter ($a,b,c$) or the 4- or
5-parameter ($a,b,c,d,e$) versions of Eqs.(\ref{eq:StrParam2},\ref%
{eq:StrParam3}). There are not enough constraints to choose among these fits.
The higher-order polynomials allow oscillatory behavior of $s^{-}(x)$ which
the 3-parameter form does not. We consider the number of crossings a
distinctive property of the physical asymmetry $s^-(x)$ rather than a
volatile function of the continuous fit parameter space. As explained in
Section \ref{sec:general}, one zero-crossing is unavoidably enforced by the
sum rule in Eq.~(\ref{eq:NumSR}). We are not aware of any solid theoretical
argument that would suggest a second crossing; nor do we find that the fits
show any preference for more than one crossing. We therefore restrict the
search for best fits in this section to one crossing.
This choice also seems to be a stable feature of
models \cite{models} based on baryon-meson fluctuations.

\item Using these candidate fits as a basis, we perform a second round of
fitting allowing the parameters associated with $s^{+}$, Eq.(\ref%
{eq:StrParam1}), to vary in addition to the $s^{-}$ variables. This improves
the fit to all data sets slightly. We observe that the shape of $s^{+}(x)$
now deviates from the starting configuration in which $s^{+}(x)$ was set
proportional to $\bar{u}(x)+\bar{d}(x)$. Defining, as in Eq.~(\ref{eq:kappa}%
), the strangeness suppression parameter $\kappa $ as the ratio of the
momentum fraction carried by the strange quarks, $[S^{+}]$, to that carried
by $\bar{u}+\bar{d}$ at $Q_{0} = 1.3 \, \mathrm{GeV}$, we find that $\kappa $
may vary in the range $0.3$\thinspace --\thinspace $0.5$: $\chi ^{2}$ has a
shallow minimum around $\kappa =0.4$. This value agrees with previous
analyses \cite{dmdata}.

Because the experimental constraints are not sufficient to uniquely
determine all the $s^{-}$ and $s^{+}$ parameters, we categorize several
classes of equally good solutions based on 
the behavior of $s^{-}(x)/s^{+}(x)$ as $x\rightarrow 0$ or $x\rightarrow 1$.

\item We finalize these classes of solutions by allowing all parton
parameters to vary so that the non-strange parton distributions can adjust
themselves to yield the best fit to all the experimental data sets. (As one
would expect, these final adjustments are generally small.) The differences
in the $\chi ^{2}$ values between the various categories of solutions are
not significant; i.e.~we find nearly degenerate minima with distinctively
different $s^-(x)$ solutions, classified according to their small $x$
behaviour. These solutions do not correspond to isolated local minima in $%
\chi^2$ space; rather they are to be thought of as specific examples of a
class of acceptable fits that lie along a nearly flat ``valley'' along which
$\chi^2$ changes very slowly.
\end{Simlis}


\subsection{Central Results}

\label{sec:results}

The quality of the fits to the global data sets other than the
CCFR-NuTeV dimuon data remains unaltered from the previous CTEQ6M analysis,
so we focus our discussion on the strangeness sector. Specifically, we
examine closely the asymmetry functions 
$s^{-}(x),\,S^{-}(x)$ and the momentum integral $[S^{-}]$. 
The asymmetry functions from three typical good fits, with different
behaviors at small $x$ (labeled as classes A,B,C), were previewed in Fig.\,%
\ref{fig:B} as illustrations.

In the accompanying table, for each sample fit we list the small-$x$
exponent $\beta_-$ [$s^-(x)\sim x^{\beta_-}$, cf.~Eq.~\ref{eq:betam}], the
integrated momentum fraction $[S^-]$, and the relative $\chi ^{2}$ values,
normalized to the $\chi ^{2}$ of solution B ($\chi^2_B$),
which we use as the reference
for comparison purposes. (Under column ``B'', we give the absolute $\chi^2$%
's in parentheses.)\footnote{%
The $\chi^2$ values of the dimuon data sets, like those of some other data
sets, do not carry rigorous statistical significance, because the correlated
systematic errors are not available and, hence, cannot be included. In the
global analysis context, the $\chi^2$ value is nevertheless used as the only
practical ``figure of merit'' for the fit. The relatively small value of the
total $\chi^2$ for the dimuon data sets, compared to the number of data
points, underlines this fact. Under this circumstance, it is common practice
to use the normalized $\chi^2$ values to compare the quality of different
fits.} To gain some insight on the constraints on the strangeness sector due
to the various types of experiments, we show separately the $\chi ^{2}$
values for the dimuon data sets, the inclusive data sets (I) that are
expected to be somewhat sensitive to $s^{-}$ (consisting of the CCFR and
CDHSW $F_3(x,Q)$ and the CDF $W$-lepton asymmetry measurements), and the
remaining data sets (II) that are only indirectly affected by $s^{-}$ (the
rest of the inclusive data sets).


\begin{center}
\begin{tabular}{|c|c||c|c|c|c|c|}
\hline
& \# pts & B+ & A & B & C & B$-$ \\ \hline\hline
$\beta_-$ & - & $-0.78$ & $-0.99$ & $-0.78$ & 0 & $-0.78$ \\ \hline
$[S^{-}]\times 100$ & - & 0.540 & 0.312 & 0.160 & 0.103 & $-0.177$ \\
\hline\hline
Dimuon & 174 & 1.30 & 1.02 & \emph{1.00} (126) & 1.01 & 1.26 \\ \hline
Inclusive I & 194 & 0.98 & 0.97 & \emph{1.00} (141) & 1.03 & 1.09 \\ \hline
Inclusive II & 2097 & 1.00 & 1.00 & \emph{1.00} (2349) & 1.00 & 1.00 \\
\hline
\end{tabular}
\\[0pt]
\rule{0em}{4ex}Table~1. The representative parton distribution sets,
arranged in order by the value of $[S^-]$.
\end{center}


Focusing on the three good fits \{A,\,B,\,C\} first, we note the following
features:

\begin{Simlis}{1em}
\item All three solutions \{A,\,B,\,C\} feature positive $[S^-]$; and the
more singular the behavior of $s^-(x)$ as $x\rightarrow 0$, the higher the
value of $[S^-]$. These are natural consequences of the strangeness sum rule
(equal $+/-$ areas under the curve of $s^-(x)$) and the small $x$
suppression of the momentum integral, as discussed earlier in Sec.~\ref%
{sec:general}.

\item Solution B is slightly favored over the other two. This, plus the fact
that its small-$x$ behavior lies in the middle of the favored range,
motivates its use as the reference fit.

\item We chose these examples among fits with the simplest parametrizations:
all cross the $x$ axis only once. With 4- or 5-parameters, which can allow
more than one crossing point, many solutions can be found that entail
oscillatory $s^-(x)$. But since the $\chi^2$ values are essentially the same
as for the simple case, we deem it premature to dwell on complicated
behaviors, which may be mere artifacts of the parametrization rather than
reflections of physical constraints. Further studies described in Sec.~\ref%
{sec:uncertainty} reinforce this point.
\end{Simlis}

\figC To show how these fits compare with data, we plot in Fig.\,\ref{fig:C}
the ratio of data/theory for the reference fit B. The four graphs correspond
to the CCFR and NuTeV neutrino and antineutrino data sets respectively. The
data points are sorted in $x$-bins, and within each $x$-bin, by $y$ value.
We see that the quality of the fit is good, within the experimental
uncertainties. There are no significant systematic deviations.
(The CCFR antineutrino data set may appear to be systematically higher than
theory. However, upon closer inspection the difference is not significant.
The data points that lie above theory consist mostly of points with large
error bars, which tend to catch the attention of the eye; whereas the fit is
actually dominated by points with small errors, which closely bracket the
theory line on both sides.
\footnote{%
This becomes apparent if the data points are re-plotted ordered by the size
of the error bars.} The value of $\chi^{2}/N$ for this data set is less than
1, comparable to those for the other sets.)

The parameters for all the fits described in this section are given in
detail in the Appendix. As is already obvious from Fig.~\ref{fig:B},
solutions with nearly degenerate $\chi^2$ may correspond to parametrizations
of $s^-(x)$ with quite different parameter values, so that a simple linear
error analysis cannot be applied. This reflects the fact that $s^-(x)$ is
not well determined as a detailed function of $x$, even when the dimuon data
are included in the fit. On the other hand, reducing the parameter space to
even fewer parameters than our minimal set would risk introducing artifacts
of an inflexible parametrization. In the next section we will, therefore,
apply the Lagrangian multiplier method to deduce the integrated momentum
asymmetry $[S^-]$ and its uncertainty.


\subsection{Range of $[S^-]$ by the Lagrange Multiplier Method}

\label{sec:range}

Beyond the best fits (A, B, C), we can study the range of $[S^-]$ consistent
with our global analysis in a quantitative way by applying the Lagrange
Multiplier (LM) method developed in \cite{LM}. By varying the Lagrange
multiplier parameter, this method explores the entire strangeness parameter
space in search of solutions with specified values of $[S^-]$,
i.e.,~constrained fits. The B$^-$ solution listed in Table 1 was obtained by
forcing $[S^-]= -0.0018$ (a relatively large negative value, but not as
large as the value $-0.0027$ quoted by \cite{NuTeV2,moch}). The B$^+$
solution was generated by forcing $[S^-]$ to go in the other (positive)
direction until the increment of the overall $\chi^2$ became comparable to
that of B$^-$; this results in $[S^-]=0.0054$.

We see from the relevant entries in Table 1 that: (i) the $\chi^2$ values of
the dimuon data sets increase by about 30\% in both B$^\pm$ fits; (ii) the
``inclusive I'' data sets disfavor the negative $[S^-]$; and (iii) the
``inclusive II'' data sets are completely neutral. These results are shown
graphically in Fig.\,\ref{fig:D}a, where the square points represent the
(relative) $\chi^2$ values of the dimuon data sets, and the triangle points
of the ``inclusive I'' data sets. (Not shown are those for the ``inclusive
II'' data sets, which remain flat (at $1.00$).) The LM fits are chosen from
a large number of fits spanning the entire strangeness parameter space. The
pattern of dependence of the $\chi^2$ values for the dimuon data sets on the
value of $[S^-]$ is nearly parabolic. This is clear evidence that the dimuon
measurement is indeed sensitive to the strangeness asymmetry as expected.
Further discussion of this observation, including the contrast to the
sensitivity of other experiments, will be given in Sec.~\ref{sec:uncertainty}%
.

\figD

We see from Fig.\,\ref{fig:D}a that, in this series of fits, the dimuon data
sets favor a range of $[S^-]$ centered around 0.0017, whereas the
``inclusive I'' data sets disfavor negative values of $[S^-]$. Fig.\,\ref%
{fig:D}b shows the dependence of the combined $\chi^2$ of the two categories
of data sets on $[S^-]$. (The $\chi^2$ of the remaining data sets used in
the global analysis are totally insensitive to $[S^-]$, cf.\,Table 1, hence
is not included in this plot.) We would like to determine a ``range of
uncertainty" of $[S^-]$ from these results. This is far from straightforward
because of well-known problems shared by all error assessments in global
analysis (mainly due to the unquantified systematic errors that show up as a
lack of statistical compatibility among the input data sets).\footnote{%
These difficulties, and practical methods to handle them, are discussed in
detail in \cite{LM,cteq6m,Mandy}.}

One naive method is to apply the $\Delta\chi^{2} = 1$ criterion.
From the
parabola in Fig.\,\ref{fig:D}b, which comes from 368 data points, this
``estimation-of-parameters criterion" corresponds to an uncertainty of $[S^-]
$ of $\pm 0.0005$. (Cf.\,the lowest horizontal line in Fig.\,\ref{fig:D}b.)
It has been known, however, that the $\Delta\chi^{2} = 1$ criterion is
unrealistic in global analysis when combining data sets with diverse
systematic errors from many different experiments \cite{LM,cteq6m,Mandy}; in
this circumstance, the overall $\chi^2$ function provides a simple measure
of relative goodness-of-fit in the minimization process, but it does not
have the strict statistical significance of a pure parameter fitting problem
as presented in textbook examples. This estimate of the uncertainty of $[S^-]
$ is far too small.

Another often-used method to evaluate "goodness-of-fit" is to apply
the {\it cumulative distribution function} $P$ for the $\chi^{2}$
distribution. One considers unacceptable values of $\chi^{2}$ greater
than $\chi^{2}_{68}$ (or $\chi^{2}_{90}$) where $P(\chi^{2} <
\chi^{2}_{f})=f$.
For 386 data points, the 68\% (90\%) criterion corresponds to $%
\chi ^{2}/\chi _{min}^{2}=1.033\ (1.1)$ respectively. These two criteria are
represented by the two upper horizontal lines in Fig.\thinspace \ref{fig:D}%
b. The uncertainty range of $[S^{-}]$ for these two cases are $\pm 0.002\
(0.003)$ respectively.

The extensive studies on quantifying uncertainties in the global analysis
context \cite{LM,cteq6m,Mandy} suggest that for this case, a realistic
range should be somewhere between the two extreme cases shown in Fig.\,\ref%
{fig:D}b. Hence we adopt the uncertainty range $0 < [S^-] < 0.004$ by this
analysis, which corresponds to the middle horizontal line in the graph (or
the $\chi^{2}_{68}$ criterion). Whereas a very small strangeness asymmetry,
consistent with zero, is not ruled out by this criterion, large negative
values of $[S^-]$ (such as -0.0027, cited in \cite{NuTeV2}) are
strongly disfavored; cf.~also Table 1. Additional sources of uncertainty
will be discussed in the following section.

\subsection{Additional Sources of Uncertainty}

\label{sec:uncertainty}

We have performed three series of studies to further assess the rebustness
of our main results. These will help us to determine a better estimate of
the overall uncertainty of  $[S^{-}]$.

\paragraph{Pure Leading Order Fits}

Since the experimental analyses of the CCFR-NuTeV dimuon data have been done
in LO QCD \cite{CN,NuTeV2}, we have carried out a whole series of purely LO
global analyses, following the same procedures as outline above, in order to
provide a basis for comparison. The results can be summarized as follows.
\begin{Simlis}{1 em}
\item The overall $\chi^2$ for the global fit increased by $\sim 200$ over
the
comparable fits described above; while the $\chi^2$'s for the dimuon data
sets actually decreased slightly.  This is not surprising, since the current
state of global analysis, with precision data from many experiments,
requires the use of NLO QCD theory -- in particular for the collider data
with typically large perturbative corrections.
On the other hand, the new dimuon data
still have comparably large experimental errors
and the NLO corrections are small
[${\cal{O}}(\lesssim 20\%)$], such that an LO fit is
adequate for them.

\item We explored the allowed range of strangeness asymmetry $[S^-]$ in
this LO study under different assumptions on the $x\rightarrow 0$ and
$x\rightarrow 1$ behavior of the $s^+(x)$ and $s^-(x)$ functions.  First, we
found that the dependence of $\chi^2_{\rm dimuon}$ on $[S^-]$ is generally
parabolic, rather similar to Fig.\,\ref{fig:D}. The
width of the distribution is comparable to Fig.\,\ref{fig:D}. The
central value for $[S^-]$ is within the range $(0,0.0015)$; the exact value
depends on the $x\rightarrow 0$ and
$x\rightarrow 1$ behavior of $s^\pm(x)$ assumed.

\item We also found that the $\chi^2_{\rm inclusive\; I}$ vs.~$[S^-]$ curve,
while generally flatter, does ``flop around'' enough for the cases studied
so that no clear pattern
can be discerned. The specific shape of this curve shown in Fig.~\ref{fig:C}
is not a common characteristic of these fits.
\end{Simlis}

\paragraph{Charm Mass Dependence}

The CCFR-NuTeV dimuon analysis treated the charm mass as one of the fit
parameters. Their analyses favored a rather high value of $m_c=1.6$\,GeV
(compared to, e.g., the PDG estimate of $1.0\,\mathrm{GeV} < m_c < 1.4\,%
\mathrm{GeV}$). The CTEQ global analyses are usually done with a fixed value
of $m_c =$ 1.3 GeV. To see whether the comparison between our results is
strongly influenced by the choice of the charm mass, we have performed
several series of fits with $m_c$ varying from 1.3 GeV to 1.7 GeV. Again,
the general features, as described above, stay the same. The central value
of $[S^-]$ does vary with the choice of $m_c$ within a given series of fits,
but the pattern is not universal. The range over which the central value
wanders is of the order $\sim 0.0015$, comparable to the width of the
parabola in Fig.~\ref{fig:D}. Unlike the specific analysis of CCFR-NuTeV,
the overall $\chi^2$ for the global analysis does favor a lower value of $%
m_c $.

\paragraph{Dependence on Decay and Fragmentation Model}

To estimate the dependence of our results on the model used to convert the
measured dimuon cross sections to structure functions for charm production,
we repeated our analyses using an alternative conversion table provided by
the CCFR-NuTeV collaboration.
\footnote{%
We thank Kevin MacFarland for supplying this table.} 
This alternative table is based on Buras-Gaemers PDFs used in CCFR-NuTeV
analyses with Collins-Spiller fragmentation functions. It is similarly tuned
to detailed features of the measured dimuon cross sections as that described
in Sec.~\ref{sec:procedure}.
\footnote{%
However, since our CTEQ6-like PDFs are rather different from the CCFR
Buras-Gaemers PDFs, it is not clear how good the approximation is to use
this conversion table. That is, the self-consistency of the procedure is not
assured.} The results obtained from the alternative fits are, again, similar
to those described earlier. The $\chi^2_{\mathrm{dimuon}}$ vs.~$[S^-]$
parabola generally has the same width as in Fig.~\ref{fig:D}. The central
value
for $[S^-]$ is in the range $0<[S^-]<0.0015$---in the lower half of the
range
quoted at the end of Sec.\,\ref{sec:range}. The dependence of
$\chi^2_{\mathrm{inclusive\; I}}$ on $[S^-]$ shows no definitive trend.

\vspace{2ex}\noindent Taken together, the results of the additional studies
described in these three paragraphs lead to several conclusions. (i) The
general features described in Secs.~\ref{sec:results} and \ref{sec:range}
are robust; (ii) The central value of $[S^-]$ wanders around within a range
that is consistent with the width of the $\chi^2_{\mathrm{dimuon}}$ vs.~$%
[S^-]$ parabola; and (iii) These additional results do not significantly
change the estimates of the previous section, except to shift the central
estimated value of $[S^-]$ to a slightly lower value, and to extend the
range of uncertainty on the lower side somewhat. The envelope of these
additional uncertainties provides an estimated range of uncertainty of the
strangeness asymmetry of\footnote{%
The uncertainties from the various sources need not be combined in
quadrature, because they are not statistically independent sources, but
rather systematic uncertainties of the theory.}
\begin{equation}
-0.001 < [S^-] < 0.004 \; .  \label{eq:range}
\end{equation}
This large range reflects both the limit of current experimental constraints
and the considerable theoretical uncertainty, as explicitly discussed in the
text. The theoretical uncertainties can be reduced in a refined NLO
analysis; the results remain to be seen. The limitations on the experimental
constraints will remain, until new experiments are done .

\section{Comparisons to previous studies}
\label{sec:compare}

A comprehensive global QCD analysis with emphasis on the strangeness sector
has been carried out previously by BPZ \cite{BPZ}.
\footnote{
As mentioned in Sec.~\ref{sec:global}, BPZ work directly with DIS cross
sections (instead of structure functions), with detailed attention to
systematic errors and other sources of uncertainties.} 
 Without the dimuon data,
which are directly sensitive to strangeness, the results of BPZ implicitly
rely on small differences between large neutral- and charged-current inclusive
cross sections from different experiments. The latest representative $s^-(x)$
and $S^-(x)$ functions extracted by 
BPZ
are shown in Fig.\,\ref{fig:A}, along with our reference
fit B. The main feature of the BPZ curves is a positive bump at rather large
$x$.
\footnote{
The BPZ curve displayed includes the more recent analysis ``with CCFR
(inclusive data).''
The original ``without CCFR'' solution in \cite{BPZ} has a more
pronounced large-x bump and a smaller negative region. An $s^-(x)$
function of such magnitude at large $x$ is in  disagreement with the
CCFR-NuTeV analysis; in addition, it would make $s(x)$ and $\bar{s}(x)$
behave quite differently 
compared with  the non-strange sea quarks and the gluon, i.e.,
$\sim (1-x)^p$, with $p$ in the range $5-10$.} 
This feature has been
attributed to the influence of the CDHSW data, particularly when re-analyzed
at the cross section level along with the other DIS experiments. Their
conclusion that data favor a positive value of the momentum integral $[S^-]$
is in general agreement with our detailed study based on the LM method.
However, the different shapes of $s^-(x)$ seen in Fig.\,\ref{fig:A} clearly
underline the difference in inputs: (i) our results are mainly dictated by the
CCFR-NuTeV dimuon data (which are not present in the BPZ analysis); (ii) their
results rely on a delicate analysis of DIS cross-section data (not
matched in our structure function analysis); and (iii) the difference in
flexibility of the parametrizations of the non-perturbative input
functions can influence the results.

The CCFR and NuTeV collaborations performed separate and combined analyses of
$s$ and $\bar{s}$ \cite{CN}, based on their own dimuon and inclusive cross
sections. To parameterize the $s(x)$ and $\bar{s}(x)$ distributions, they
chose the model formula
\begin{equation}\label{eq:CN}
\left(
\begin{array}{c}
s(x,Q) \\
\bar{s}(x,Q)%
\end{array}%
\right) = \frac{\bar{u}(x,Q)+\bar{d}(x,Q)}{2}\left(
\begin{array}{c}
\kappa (1-x)^{\alpha } \\
\bar{\kappa}(1-x)^{\bar{\alpha}}%
\end{array}%
\right)
\end{equation}
for all $\{x,Q\}$, where $\kappa, \bar{\kappa}, \alpha, \bar{\alpha}$ are
fitting parameters.

Curves representing the general behavior of the model (\ref{eq:CN}) at
$Q^{2}=10$\,GeV$^{2}$, with parameter values ($\kappa ,\bar{\kappa},\alpha
,\bar{\alpha}$) taken from \cite{CN}, have been shown in Fig.\,\ref{fig:A} for
comparison with the other distributions. While this model might be acceptable for a
limited range of $x$ and $Q$, it leads to problems when extrapolated to general 
$\{x,Q\}$ values: (i) the
strangeness number sum rule $[s^-]=0$ is violated (in fact, the
integral $[s^-]$ diverges unless $\kappa=\bar{\kappa}$), as is the
momentum sum rule, Eq.~\ref{eq:MomSR}; (ii) the QCD evolution
equations are violated  by the $Q$-dependent
paramaterizations of the PDFs.\footnote{These problems are  independent of the specific
issues of the strange quark asymmetry.} %
The first problem is evident in Fig.\,\ref{fig:A}.
\footnote{Re-analysis of the CCFR-NuTeV data by the experimental group,
taking into account these issues, are underway.  Initial results from
partial implementations of the above-mentioned theoretical constraints were
reported by P.~Spentzouris at \emph{International Workshop on Weak
Interactions and Neutrinos 2003} (WIN03), Lake Geneva, October, 2003.}

It could be argued that, since the experiment only covers a limited
range of $x$, the enforcement of the sum rules is not critical in
extracting limited information on $s(x,Q)$ and $\bar{s}(x,Q)$.  Given
this,the CCFR curve in Fig.~\ref{fig:A} implies negative $s^-(x)$ over
most of the experimental $x$ range (0.01 -- 0.3).
\footnote{This is consistent with the fact that (without the constraint of sum
rules) \cite{NuTeV2} quotes $[S-]=-0.0027\pm 0.0013$.} %
The uncertainty is smaller at the lower $x$ end because of better statistics.
It is this feature of the data that we invoked in the general discussion of
Sec.~\ref{sec:general}.

To describe the behavior of $s^-(x)$ over the full $x$ range, the
strangeness number sum rule must be enforced.  The sum rule also provides a powerful
theoretical constraint on the data analysis, as demonstrated in
Secs.~\ref{sec:general} and \ref{sec:global}. 
For functions to be candidate  \emph{universal} parton distributions of the PQCD
formalism, they must satisfy both the sum rules and  the QCD evolution equations.

\section{Conclusion}
 
We find a range of solutions in the strangeness sector that are
consistent with all relevant world data used in the global analysis. The
dimuon data are vital in constraining the strangeness asymmetry parameters.
The constraints provided by other inclusive measurements, labeled as
``inclusive I'' in the text, are consistent with those provided by dimuon
data, although much weaker. The
allowed solutions generally prefer the momentum integral $[S^{-}]\equiv
\int_{0}^{1}\,x[s(x)-\bar{s}(x)]\ dx$ to be positive. This conclusion is quite
robust, and it follows  from the basic properties of PQCD and from qualitative
features of the experimental data. However, the size of this strangeness
momentum asymmetry is still quite uncertain; we can only estimate that
$\left[S^{-}\right]$ lies in the range from $-0.001$ to $+0.004$. The Lagrange
Multiplier method explicitly demonstrates that both the dimuon data and the
``inclusive I'' data sets strongly disfavor a large negative value of $[S^-]$,
although they may still be consistent with zero asymmetry.

The fact that $\left[S^{-}\right]$ has a large uncertainty has significant
implications for the precision measurement of the weak mixing angle,
$\sin^{2}\theta_W$, from neutrino scattering. This issue is studied separately
in Ref.~\cite{paper2}.

This paper marks the first global QCD analysis incorporating direct
experimental constraints on the strangeness sector.  We have so far focused
only on the strangeness asymmetry, which represents a new frontier in
parton degrees of freedom.  Much still needs to be done to improve the
treatment of the dimuon data (to true NLO accuracy), and to fully explore
all the degrees of freedom associated with $s^+(x)$ and $s^-(x)$. As
progress is made on these fronts, the uncertainty on $[S^-]$ will no doubt
decrease as well.  

\noindent\textbf{Note added:}
After this manuscript was completed an investigation of the 3-loop
perturbative strangeness asymmetry was presented in 
\cite{Catani}.

\vspace{5ex}
\noindent\textbf{Acknowledgment}

We thank members of the CCFR and NuTeV collaboration, in particular
T.\,Bolton and M.\,Goncharov, for discussions about the dimuon data, and
assistance in their use, and Kevin MacFarland, Dave Mason, and  Panagiotis Spentzouris
for useful comments and suggestions. We also thank Benjamin Portheault for
discussions of the BPZ work and for assistance in generating their results
for comparison. F.O. acknowledges the hospitality of MSU and BNL where a
portion of this work was performed. 
This research was supported by the
National Science Foundation (grant No.~0100677),  RIKEN, Brookhaven
National Laboratory, and the U.S. Department of Energy (Contract
No.~DE-AC02-98CH10886, and No.~DE-FG03-95ER40908), and by the Lightner-Sams
Foundation.



\section*{Appendix}

In this appendix, we provide some detailed information on the
parametrization of the non-perturbative parton distribution functions used
in our global analysis work, as well as the values of the parameters for the
representative fits.

The non-strange parton distributions at $Q=Q_{0}=1.3$ GeV are parametrized
as in CTEQ6. We use
\begin{equation}
x\,f(x,Q_{0})=A_{0}\,x^{A_{1}}\,(1-x)^{A_{2}}\,e^{A_{3}x}%
\,(1+e^{A_{4}}x)^{A_{5}}  \label{eq:param}
\end{equation}%
with independent parameters for parton flavor combinations $u_{v}\equiv u-%
\bar{u}$, $d_{v}\equiv d-\bar{d}$, $g$, and $\bar{u}+\bar{d}\,$. \ To
distinguish the $\bar{d}$ and $\bar{u}$ distributions, we parametrize the
\emph{ratio} $\bar{d}/\bar{u}$, as a sum of two terms:
\begin{equation}
\bar{d}(x,Q_{0})/\bar{u}(x,Q_{0})=A_{1}\,x^{A_{2}}\,(1-x)^{A_{3}}\;+%
\;(1+A_{4}\,x)\,(1-x)^{A_{5}}\;.  \label{eq:dou}
\end{equation}

\bigskip The strangeness sector is parametrized according to the description
of Sec.(\ref{sec:param}) with%
\begin{equation*}
xs^{+}(x,Q_{0})=A_{0}\,x^{A_{1}}\,(1-x)^{A_{2}}\,e^{A_{3}x}%
\,(1+e^{A_{4}}x)^{A_{5}}
\end{equation*}%
where the $A_{1,2,3,4,5}$ coefficients are either equated to those of $\bar{u%
}+\bar{d},$ or allowed to vary independently, depending on the particular
fit being performed; and%
\begin{equation*}
s^{-}(x,Q_{0})=s^{+}(x,Q_{0})\tanh [A_{0}\,x^{A_{1}}(1-x)^{A_{2}}\left( 1-%
\frac{x}{A_{3}}\right) \left( 1+A_{4}~x+A_{5}~x^{2}\right) ]
\end{equation*}

As an example, the ``standard fit'' (B-fit) described in Sec.\ref{sec:results}
has the following coefficients.

\centerline{
\begin{tabular}{|c|c|c|c|c|c|c|}
\hline
    B-fit         & $A_{0}$ & $A_{1}$ & $A_{2}$ & $A_{3}$ & $A_{4}$ & $A_{5}$ \\ \hline
$d_{v}$           & 1.55891 & 0.62792 & 5.08865 & 0.70688 & -0.26338 & 3.00000   \\ \hline
$u_{v}$           & 1.73352 & 0.55260 & 2.90090 & -2.63846 & 1.45774 & 1.85987   \\ \hline
$g$               & 31.24912 & 0.52014 & 2.38230 & 4.23010 & 2.33765 & -3.00000  \\ \hline
$\bar{d}/\bar{u}$ & -- & 10.23811 & 5.19599 & 14.85860 & 17.00000 & 8.68941      \\ \hline
$\bar{u}+\bar{d}$ & 0.06758 & -0.29681 & 7.71700 & -0.82223 & 4.43481 & 0.66711  \\ \hline
$s+\bar{s}$       & 0.04966 & 0.03510 & 7.44149 & -1.44570 & 5.13400 & 0.59659   \\ \hline
$s-\bar{s}$       & -0.08438 & 0.18803 & 2.36708 & 0.04590 & 0.00000 & 0.00000   \\ \hline
\end{tabular}
}

\medskip\noindent And the other sample fits have coefficients:

\begin{center}
\begin{tabular}{|c|c|c|c|c|c|c|}
\hline
A-fit & $A_{0}$ & $A_{1}$ & $A_{2}$ & $A_{3}$ & $A_{4}$ & $A_{5}$ \\ \hline
$d_{v}$ & 1.55485 & 0.62740 & 5.09550 & 0.70010 & -0.25130 & 3.00000 \\
\hline
$u_{v}$ & 1.73345 & 0.55260 & 2.90090 & -2.63740 & 1.45810 & 1.85910 \\
\hline
$g$ & 31.05518 & 0.51840 & 2.38230 & 4.24730 & -2.33790 & 3.00000 \\ \hline
$\bar{d}/\bar{u}$ & -- & 10.23920 & 5.19690 & 14.85860 & 17.00000 & 8.68530
\\ \hline
$\bar{u}+\bar{d}$ & 0.06677 & -0.29810 & 7.71700 & -0.80540 & 4.45630 &
0.66420 \\ \hline
$s+\bar{s}$ & 0.05133 & 0.06567 & 7.59880 & -1.43972 & 5.13400 & 0.61715 \\
\hline $s-\bar{s}$ & -0.02000 & -0.06562 & -0.48403 & 0.01987 & -0.58837 &
0.00000 \\ \hline
\end{tabular}

\vspace{3ex} 
\begin{tabular}{|c|c|c|c|c|c|c|}
\hline
C-fit & $A_{0}$ & $A_{1}$ & $A_{2}$ & $A_{3}$ & $A_{4}$ & $A_{5}$ \\ \hline
$d_{v}$ & 1.39587 & 0.60594 & 4.75379 & -0.75291 & 0.31387 & 3.00000 \\
\hline
$u_{v}$ & 1.72080 & 0.55260 & 2.90090 & -2.36331 & 1.60950 & 1.59812 \\
\hline
$g$ & 29.40547 & 0.50423 & 2.38230 & 4.41409 & 2.34266 & -3.00000 \\ \hline
$\bar{d}/\bar{u}$ & -- & 10.23202 & 5.19618 & 14.85860 & 17.00000 & 8.62620
\\ \hline
$\bar{u}+\bar{d}$ & 0.06479 & -0.30038 & 7.71700 & -0.68604 & 4.61325 &
0.62988 \\ \hline
$s+\bar{s}$ & 0.04289 & 0.00809 & 7.71700 & -1.22365 & 5.13400 & 0.62988 \\
\hline
$s-\bar{s}$ & -2.36926 & 0.99191 & 8.23499 & 0.07884 & 0.00000 & 0.00000 \\
\hline
\end{tabular}

\vspace{3ex} 
\begin{tabular}{|c|c|c|c|c|c|c|}
\hline
B$_+$-fit & $A_{0}$ & $A_{1}$ & $A_{2}$ & $A_{3}$ & $A_{4}$ & $A_{5}$ \\
\hline
$d_{v}$ & 1.43611 & 0.61170 & 4.73270 & -0.67420 & 0.24920 & 3.00000 \\
\hline
$u_{v}$ & 1.71921 & 0.55260 & 2.90090 & -2.39220 & 1.60430 & 1.61490 \\
\hline
$g$ & 29.76781 & 0.50800 & 2.38230 & 4.35570 & -2.33860 & 3.00000 \\ \hline
$\bar{d}/\bar{u}$ & -- & 10.19560 & 5.16810 & 14.85860 & 17.00000 & 8.69760
\\ \hline
$\bar{u}+\bar{d}$ & 0.06729 & -0.29650 & 7.71700 & -0.75670 & 4.52290 &
0.64380 \\ \hline
$s+\bar{s}$ & 0.03456 & 0.00210 & 8.23420 & -1.26970 & 5.13410 & 0.72501 \\
\hline $s-\bar{s}$ & -0.40480 & 0.22103 & 3.40190 & 0.04701 & 0.31550 &
0.00000 \\ \hline
\end{tabular}

\vspace{3ex} 
\begin{tabular}{|c|c|c|c|c|c|c|}
\hline
B$_-$-fit & $A_{0}$ & $A_{1}$ & $A_{2}$ & $A_{3}$ & $A_{4}$ & $A_{5}$ \\
\hline
$d_{v}$ & 1.43611 & 0.61170 & 4.73270 & -0.67420 & 0.24920 & 3.00000 \\
\hline
$u_{v}$ & 1.71921 & 0.55260 & 2.90090 & -2.39220 & 1.60430 & 1.61490 \\
\hline
$g$ & 29.76781 & 0.50800 & 2.38230 & 4.35570 & -2.33860 & 3.00000 \\ \hline
$\bar{d}/\bar{u}$ & -- & 10.19560 & 5.16810 & 14.85860 & 17.00000 & 8.69760
\\ \hline
$\bar{u}+\bar{d}$ & 0.06717 & -0.29650 & 7.71700 & -0.75670 & 4.52290 &
0.64380 \\ \hline
$s+\bar{s}$ & 0.04356 & 0.00210 & 7.33918 & -1.26970 & 5.13400 & 0.60083 \\
\hline $s-\bar{s}$ & 0.01781 & 0.22103 & -15.02691 & 0.22693 & -1.23666 &
0.00000 \\ \hline
\end{tabular}
\end{center}

\end{document}